\documentclass[epj]{svjour}
\usepackage{amsfonts,amssymb,amsmath}
\usepackage{graphicx,color}
\usepackage{dcolumn,slashed}

\makeatletter
\DeclareRobustCommand{\cev}[1]{
  {\mathpalette\do@cev{#1}}
}
\newcommand{\do@cev}[2]{
  \vbox{\offinterlineskip
    \sbox\z@{$\m@th#1 x$}
    \ialign{##\cr
      \hidewidth\reflectbox{$\m@th#1\vec{}\mkern4mu$}\hidewidth\cr
      \noalign{\kern-\ht\z@}
      $\m@th#1#2$\cr
    }
  }
}
\makeatother

\usepackage{color}
\usepackage{listings}
\usepackage[utf8]{inputenc}
\usepackage[english]{babel}
\usepackage{graphicx}
\usepackage{physics}
\usepackage{amsmath}
\usepackage{amssymb}
\usepackage{bbm}
\usepackage{hyperref}
\usepackage{tikz}
\usepackage{xcolor}
\usepackage{mathtools}
\usepackage{slashed}
\usepackage{soul}
\newcommand{\K}[1]{\ensuremath{\left(#1\right)}}
\newcommand{\Ke}[1]{\ensuremath{\left[#1\right]}}

\newcommand{\vn}{\ensuremath{\boldsymbol{n}}}

\allowdisplaybreaks

\begin{document}

%
%\title{Insert your title here\thanksref{t1}}

\title{Lattice Monte Carlo Simulations with Two Impurity Worldlines} 

{\color{red}
\author{Fabian Hildenbrand \inst{1} 
\and Serdar Elhatisari \inst{2,3} 
\and Timo A. L\"{a}hde \inst{1} 
\and Dean Lee\inst{4}  
\and Ulf-G.~Mei{\ss}ner\inst{3,1,5}
%
%First author\inst{1} \and Second author\inst{2}% etc
%% \thanks is optional - remove next line if not needed
}                     % Do not remove
\institute{
Institut~f\"{u}r~Kernphysik,~Institute~for~Advanced~Simulation and
J\"{u}lich~Center~for~Hadron~Physics, Forschungszentrum~J\"{u}lich, \\D-52425~J\"{u}lich,~Germany 
\and Faculty of Natural Sciences and Engineering, Gaziantep Islam Science and
Technology University,\\ Gaziantep 27010, T\"{u}rkiye
\and Helmholtz-Institut~f\"{u}r~Strahlen-~und~Kernphysik~and~Bethe~Center~for
Theoretical~Physics, Universit\"{a}t~Bonn, \\ D-53115~Bonn,~Germany
\and Facility for Rare Isotope Beams and Department of Physics and Astronomy, Michigan State University, \\ 
MI 48824, USA
\and Tbilisi State University, 0186 Tbilisi, Georgia
}
}

\date{Received: date / Revised version: date}

\abstract{
%In this paper, 
We develop the impurity lattice Monte Carlo formalism, for the case of two distinguishable impurities in a bath of polarized fermions.
The majority particles are treated as explicit degrees of freedom, while the impurities are described by worldlines. The latter serve as localized auxiliary fields, which affect the majority particles. 
We apply the method to non-relativistic three-dimensional systems of two impurities and a number of majority particles where both the impurity-impurity interaction and the impurity-majority 
interaction have zero range. We consider the case of an attractive impurity-majority interaction, and we study the formation and disintegration of bound states as a function of the 
impurity-impurity interaction strength. We also discuss the potential applications of this formalism to other quantum many-body systems.
 % end of PACS codes
} %end of abstract

\maketitle

%%%%%%%%%%%%%%%%%%%%%%%%%%%%%%%%%%%%%%%%%%%%%%

\section{Introduction}

There is growing interest in the study of nuclear processes and phenomena including
hyperons, as these have important influences on the properties of finite nuclei and infinite
nuclear matter~\cite{Petschauer:2020urh,Tolos:2020aln,Vidana:2021ppe,Hildenbrand:2020kzu}.
The development of an {\it ab initio} formalism to study hypernuclear physics is an
important step along this direction. The auxiliary-field quantum Monte Carlo (AFQMC) method is particularly well-suited to
study the properties of finite nuclei and nuclear matter. This holds in particular for Wigner's SU(4)
symmetry~\cite{Wigner:1936dx} where the interaction is independent of spin and isospin. Then,
the AFQMC method becomes very powerful as it is free of the fermion sign problem~\cite{Lu:2018bat,Lu:2019nbg}.
However, when hyperons are included in AFQMC simulations, the sign problem resurfaces since
there is no longer an approximate symmetry for the interactions~\cite{bour2009}. Therefore,
hypernuclear physics requires an alternative approach such as the so-called the impurity
lattice Monte Carlo (ILMC) method, which is the subject of this paper. 

ILMC is a ``hybrid'' algorithm, in the sense that it combines AFQMC with worldline MC simulations, 
and provides a powerful means to study non-relativistic, strongly coupled systems, such as polarons in various dimensions, or hypernuclei.  
The basic assumption is that the minority species of impurities, such as hyperons
in a nucleus, is treated differently than the majority species of particles, which are filling the background in a medium.
In the case of hypernuclei, the majority particles are nucleons. The ILMC method has been introduced in
Ref.~\cite{Elhatisari:2014lka} in the context of a Hamiltonian 
theory of spin-up and spin-down fermions, and applied to the intrinsically
non-perturbative physics of Fermi polarons in two dimensions in Ref.~\cite{Bour:2014bxa}.

The ILMC method has proven to be successful for the case in which only one fermion (of a given species)
is immersed in a sea of particles from one or more other species. ILMC leads to
a formalism in which only the majority species fermions appear as explicit degrees of freedom, while the minority fermion 
is represented by a worldline in Euclidean projection time. The spatial position of
this worldline is updated using Metropolis moves, while the interactions between
the majority fermions are described by auxiliary fields.
The efficient performance of ILMC was explicitly demonstrated in Ref.~\cite{Bour:2014bxa} for a calculation with
10 spin-up fermions (the sea of majority particles) and one spin-down impurity.
The first application of ILMC to the inclusion of hyperons into
Nuclear Lattice Effective Field Theory (NLEFT) simulations~\cite{Lee:2008fa,Lahde:2019npb}
was performed in Ref.~\cite{Frame:2020mvv}. In that work, the $\Lambda$ hyperon
was considered as the minority species and represented by a worldline in Euclidean time. This
$\Lambda$ worldline was then immersed in a medium consisting of an arbitrary number
of nucleons $N$. With simple spin- and isospin-independent hyperon-nucleon
and nucleon-nucleon interactions, the binding energies of the hypertriton, $^4_\Lambda$H/He, and  $^5_\Lambda$He were calculated, leading
to a qualitative agreement with experiment. Most importantly, the computational effort for ILMC was found
to scale approximately linearly with the number of nucleons. This paves the way for the
calculation of light, medium-mass and heavy hypernuclei.

Here, we extend ILMC to the case of 
two interacting, distinguishable impurities represented by worldlines.
Besides applications in atomic physics, this extension is necessary
if one wants to consider double-$\Lambda$ hypernuclei within NLEFT,
see~\cite{Gal:2016boi,Hiyama:2018lgs} for recent reviews. In this work, we consider two distinguishable impurities 
immersed in a sea of polarized fermions of the same species. For convenience of notation, we refer to the two impurities 
as spin-up fermions ($\uparrow_a$ and $\uparrow_b$), while the majority background particles will be labeled as spin-down fermions, $\downarrow$.  

For simplicity, we take the masses of all particles to be the same. We take the interactions between each impurity and the background particles to be a zero-range 
attractive interaction with the same strength for $\uparrow_a+\downarrow$ and $\uparrow_b+\downarrow$. We take the impurity-impurity interaction $\uparrow_a+\uparrow_b$ to 
also have zero range, but we consider both the attractive and repulsive cases. Our aim is to calculate the binding energies of these $N$-body systems and draw some
conclusions for the fate of double-$\Lambda$ hypernuclei from such a simplified scenario.

This paper is organized as follows: In Sec.~\ref{sec:Ham}, we present the Hamiltonian
used for this study. In Sec.~\ref{sec:two}, we extend the impurity worldline formalism
to the case of two impurities, and focus mostly on the differences to the
one-worldline case developed and described in Ref.~\cite{Elhatisari:2014lka}.
In Sec.~\ref{sec:res}, we present our results for the binding energies of the systems
made from two impurities and an arbitrary number of background particles.
In Sec.~\ref{sec:disc}, we conclude by a discussion of
future directions and applications of the impurity Monte Carlo
method to the problem of hypernuclei in NLEFT. In the Appendix, we benchmark
our approach by calculating the triton binding energy for an
interaction that is independent of spin and isospin.

%%%%%%%%%%%%%%%%%%%%%%%%%%%%%%%%%%%%%%%%%%%%%%

\section{Lattice Hamiltonian}
\label{sec:Ham}

%%%%%%%%%%%%%%%%%%%%%%%%%%%%%%%%%%%%%%%%%%%%%%

We consider systems of two distinguishable impurity particles
labeled $a$ and $b$, interacting with a background sea of indistinguishable spin-down fermions. For simplicity, 
we assume that impurities and spin-down fermions have equal mass $m$. A lattice Hamiltonian can then be constructed in the following way. 
We start with the non-relativistic Hamiltonian of the free theory,
\begin{align}
\hat{H}_0 = \frac{1}{2m}\sum_{s=\uparrow_a,\uparrow_b,\downarrow}\int\dd[3]{\boldsymbol{r}}
\boldsymbol{\nabla} a_s^\dagger\K{\boldsymbol{r}}\boldsymbol{\nabla}a_s^{}\K{\boldsymbol{r}},
\end{align}
in which $a_s$ and $a_s^\dagger$ represent annihilation and creation operators,
and $i \in \mathbb{N}$ is the labeling index for the background particles. The zero-range interaction 
between the impurities and background particles is controlled by the coupling constant $C_{IB}$. This corresponds to the
low-energy limit of a theory in which the scattering length $a_{\text{scatt}}$ is large compared
to the effective range $r_{\text{eff}}$. In a similar fashion, we introduce a contact interaction
between the two impurities $C_{II}$. The interaction Hamiltonian is given by
\begin{align}
\hat{H}_I = \: & C_{II}\int\dd[3]{\boldsymbol{r}}\hat{\rho}_{\uparrow_b}\K{\boldsymbol{r}}\hat{\rho}_{\uparrow_a}
\K{\boldsymbol{r}} \nonumber \\
& + C_{IB}\int\dd[3]{\boldsymbol{r}}\, \bigg[ \hat{\rho}_{\uparrow_a}\K{\boldsymbol{r}}
\hat{\rho}_{\downarrow}\K{\boldsymbol{r}}
+\hat{\rho}_{\uparrow_b}\K{\boldsymbol{r}}\hat{\rho}_{\downarrow}
\K{\boldsymbol{r}}\bigg],
\end{align}
where $\hat{\rho}_{s}\K{\boldsymbol{r}}$ are density operators given by
\begin{align}
\hat{\rho}_{s}\K{\boldsymbol{r}}= a_s^\dagger\K{\boldsymbol{r}}a_s^{}\K{\boldsymbol{r}}.
\end{align}
As in all lattice calculations, the ultraviolet physics of the zero-range interactions are
regulated by the lattice spacing. We denote the spatial lattice spacing by $a$, and the temporal
lattice spacing (due to the Trotter decomposition of the Euclidean time evolution)
by $a_t$. We express all physical quantities 
in lattice units by multiplying them with corresponding powers of $a$, so as to form dimensionless combinations. 
We also define the lattice spacing ratio $\alpha = a_t/a$.
The free lattice Hamiltonian is
\begin{align}
\hat{H}_0=\hat{H}_0^{\uparrow_a}+\hat{H}_0^{\uparrow_b}+\hat{H}_0^{\downarrow},
\end{align}
with 
\begin{align}
%\begin{split}
H_0^s = \: & \frac{1}{2m}\sum_{\vn}\sum_{i=1}^3a^\dagger_s\K{\vn}
\nonumber \\
 &\times\Ke{2a_s\K{\vn}-a_s\K{\vn+\hat{l}_i}-a_s\K{\vn-\hat{l}_i}},
% \end{split}
\end{align}
where the $\hat{l}_i, i\in {1,2,3}$ are unit vectors in the spatial dimensions. 
We express the various lattice interaction terms in a compact form, as
\begin{align}
H_{s's} = C_{s's}\sum_{\vn}\hat{\rho}_{s'}\K{\vn}\hat{\rho}_{s}\K{\vn},
\end{align}
where the couplings $C_{s's}$ can be tuned
in such a way as to either produce specific dimer binding energies, or scattering lengths
via L\"uscher's formula~\cite{Luscher:1986pf,Luscher:1990ux}.

\section{Impurity Lattice Monte Carlo: Extension to Two Impurities \label{sec:two}}

The ILMC method for a single impurity has already been treated in the literature~\cite{Elhatisari:2014lka,Bour:2014bxa,Frame:2020mvv}. 
We now extend the formalism to systems with two impurities. As in the single impurity case, the goal is to integrate out all impurities from the lattice action 
and to reduce the explicit degrees of freedom in the Monte Carlo simulation. We also derive the transfer matrix formalism for two impurities in a sea of an arbitrary number 
of spin-down particles. This can be achieved using the exact correspondence between the Grassmann path integral and the normal-ordered 
transfer matrix formalisms~\cite{Lahde:2019npb}.

We define our system in a three-dimensional periodic cubic box of length $L$ in the spatial directions and $L_{t}$ in the temporal direction. 
We denote anti-commuting Grassmann variables by $\theta^{\,}_{s}$ and $\theta^{*}_{s}$ and choose the Grassmann variables to be periodic 
in the spatial directions and anti-periodic in the temporal direction. The path integral formula for the Grassmann variables is
\begin{align}
\mathcal{Z} = \int
\bigg[
\prod_{\substack{n_{t}, \boldsymbol{n} \\ s = {\uparrow_a},{\uparrow_b},\downarrow}}
d\theta^{\,}_{s}\left(n_{t},\boldsymbol{n}\right)
d\theta^{*}_{s}\left(n_{t},\boldsymbol{n}\right)
\bigg]
e^{-S\left[\theta,\theta^*\right]},
\label{eqn:Grassmann-path-int}
\end{align}
where $S\left[\theta,\theta^*\right]$ is the lattice action in terms of the Grassmann variables, given by
\begin{align}
S\left[\theta,\theta^*\right]
= & \sum_{n_{t}}\bigg\{
S_{t}\left[\theta,\theta^*,n_{t}\right] + S_{H_0}\left[\theta,\theta^*,n_{t}\right]
\nonumber \\
& + S_{H_{SI}}\left[\theta,\theta^*,n_{t}\right] + S_{H_{II}}\left[\theta,\theta^*,n_{t}\right]
\bigg\},
\end{align}
with
\begin{align}
S_{t}\left[\theta,\theta^*,n_{t}\right]
= & \!\! \sum_{s = {\uparrow_a},{\uparrow_b},\downarrow}
\sum_{\boldsymbol{n}}
\left[\theta^{*}_{s}{\left(n_{t}+\hat{0},\boldsymbol{n}\right)}
-\theta^{*}_s\left(n_t,\boldsymbol{n}\right)\right]
\nonumber\\
&\times \theta^{\,}_s\left(n_t,\boldsymbol{n}\right)
\,,
\end{align}
and
\begin{align}
& S_{H_0} \left[\theta,\theta^*,n_{t}\right]
= W_h \!\!
\sum_{s = {\uparrow_a},{\uparrow_b},\downarrow}
\sum_{\boldsymbol{n}}\sum_{i=1}^3\theta^*_s{\left(n_{t},\boldsymbol{n}\right)}
\nonumber\\
&\quad \times \bigg[2\theta^{\,}_{s}{\left(n_{t},\boldsymbol{n}\right)}
-\theta^{\,}_s\left(n_t,\boldsymbol{n}+\hat{l}_i\right)-\theta^{\,}_s\left(n_t,\boldsymbol{n}-\hat{l}_i\right)\bigg],
\end{align}
where $W_h$ is given by Eq.~(\ref{eqn:Wh}). For the interaction terms, we find
\begin{align}
S_{H_{SI}}\left[\theta,\theta^*,n_{t}\right] = & \: \alpha \, C_{IB}
\sum_{\boldsymbol{n}}
\sum_{s = \uparrow_a,\uparrow_b}
\theta^*_s{\left(n_{t},\boldsymbol{n}\right)}
\theta^{\,}_s{\left(n_{t},\boldsymbol{n}\right)}
\nonumber\\
&\times\theta^*_{\downarrow}{\left(n_{t},\boldsymbol{n}\right)}
\theta^{\,}_{\downarrow}{\left(n_{t},\boldsymbol{n}\right)},
\end{align}
and
\begin{align}
S_{H_{II}}\left[\theta,\theta^*,n_{t}\right] = & \: \alpha \, C_{II}
\sum_{\boldsymbol{n}}
\theta^*_{\uparrow_a}{\left(n_{t},\boldsymbol{n}\right)}
\theta^{\,}_{\uparrow_a}{\left(n_{t},\boldsymbol{n}\right)}
\nonumber\\
&\times\theta^*_{\uparrow_b}{\left(n_{t},\boldsymbol{n}\right)}
\theta^{\,}_{\uparrow_b}{\left(n_{t},\boldsymbol{n}\right)}.  
\end{align}

While Eq.~(\ref{eqn:Grassmann-path-int}) is convenient for deriving lattice Feynman rules, the transfer matrix operator formalism is more amenable to lattice MC simulations. 
For a detailed derivation of the exact relation between the Grassmann path integral formula and the normal-ordered transfer matrix formalism, we refer the reader to 
Refs.~\cite{Lee:2008fa,Lahde:2019npb}. We can write the amplitude $\mathcal{Z}$ as
\begin{align}
\mathcal{Z} = \tr \, \hat{M}^{L_{t}}
\,,
\end{align}
where $\hat{M}$ is the normal-ordered transfer matrix.

We start from the occupation number
basis~\cite{Elhatisari:2014lka}. With $\chi^s_{n_t}\K{\vn}$ counting the occupation number on each
lattice site for any particle species, we can write any configuration as
\begin{align}
&\ket{\chi_{n_{t}}^\downarrow,\chi_{n_{t}}^{\uparrow_a},\chi_{n_{t}}^{\uparrow_b}} =
\nonumber\\
&\prod_{\vn}\Ke{a_\downarrow^\dagger\K{\vn}}^{\chi_{n_t}^\downarrow\K{\vn}}\Ke{a_{\uparrow_a}^\dagger\K{\vn}}^{\chi_{n_t}^{\uparrow_a}
\K{\vn}}\Ke{a_{\uparrow_b}^\dagger\K{\vn}}^{\chi_{n_t}^{\uparrow_b}\K{\vn}}\!\ket{0}.
%&&
\end{align}
Without loss of generality, we can take all of the particles to be fermions, including the two distinguishable impurities. Thus the occupation numbers can 
assume the values~$0$ or~$1$. We shall use this property to determine the transfer matrix elements between two successive time steps $n_t$ and $n_{t+1}$ 
from the lattice Grassmann functions, 
\begin{align}
\begin{split}
&\matrixel{\chi_{n_{t+1}}^{\downarrow},\chi_{n_{t+1}}^{\uparrow_a},\chi_{n_{t+1}}^{\uparrow_b}}
{ \, \hat{M}\, }
{\chi_{n_{t}}^{\downarrow},\chi_{n_{t}}^{\uparrow_a},\chi_{n_{t}}^{\uparrow_b}}=\\
&\prod_{\boldsymbol{n}}\left(\left[\frac{\overrightarrow{\partial}}{\partial\theta_{\uparrow_b}^*{\left(n_{t},\boldsymbol{n}\right)}}\right]^{\chi_{n_{t+1}}^{\uparrow_b}\left(\boldsymbol{n}\right)}
\left[\frac{\overrightarrow{\partial}}{\partial\theta_{\uparrow_a}^*{\left(n_{t},\boldsymbol{n}\right)}}\right]^{\chi_{n_{t+1}}^{\uparrow_a}\left(\boldsymbol{n}\right)}\right.\\
&\left.\times\left[{\frac{\overrightarrow{\partial}}{\partial\theta_\downarrow^*{\left(n_{t},\boldsymbol{n}\right)}}}\right]^{\chi_{n_{t+1}}^{\downarrow}\left(\boldsymbol{n}\right)}\right)
\times X\left(\boldsymbol{n}\right)M\left(n_t\right)\\
&\times\prod_{\boldsymbol{n}'}\left(\Ke{\frac{\overleftarrow{\partial}}{\partial\theta_{\downarrow}{\left(n_{t},\boldsymbol{n}'\right)}}}^{\chi_{n_{t}}^{\downarrow}\left(\boldsymbol{n}'\right)}\left[{\frac{\overleftarrow{\partial}}{\partial\theta_{\uparrow_a}{\left(n_{t},\boldsymbol{n}'\right)}}}\right]^{\chi_{n_{t}}^{\uparrow_a}\left(\boldsymbol{n'}\right)}\right.\\
&\left.\left.\times\left[{\frac{\overleftarrow{\partial}}{\partial\theta_{\uparrow_b}{\left(n_{t},\boldsymbol{n}'\right)}}}\right]^{\chi_{n_{t}}^{\uparrow_b}\left(\boldsymbol{n}'\right)}\right)\right |_{\begin{subarray}{c}
\theta_\downarrow=\theta_{\uparrow_a}=\theta_{\uparrow_b}=0 \\
\theta^*_\downarrow=\theta^*_{\uparrow_a}=\theta^*_{\uparrow_b}=0 \\
\end{subarray}} \,,
\end{split}
\end{align}
where the $::$ represent the normal ordering of operators. 
Here, $X\K{\vn}$ and $M\K{n_t}$ are the Grassmann functions
\begin{align}
X\K{\vn} = & \prod_{\vn} e^{\theta^*_\downarrow\Ke{n_t,\vn}\theta_\downarrow\Ke{n_t,\vn}}
e^{\theta^*_{\uparrow_a}\Ke{n_t,\vn}\theta_{\uparrow_a}\Ke{n_t,\vn}}\nonumber\\
 & \times e^{\theta^*_{\uparrow_b}\Ke{n_t,\vn}\theta_{\uparrow_b}\Ke{n_t,\vn}},
\end{align}
and
\begin{align}
M\K{n_t} = \exp\K{-S_{H_0}}\exp\K{-S_{H_{SI}}}\exp\K{-S_{H_{II}}}.
\end{align}
The above matrix elements of the transfer matrix operator are non-zero only if
\begin{eqnarray}
\sum_{\boldsymbol{n}}\chi_{n_{t}}^{\uparrow_a}(\boldsymbol{n})
= \sum_{\boldsymbol{n}} \chi_{n_{t+1}}^{\uparrow_a}(\boldsymbol{n}) = 1,
\end{eqnarray}
and 
\begin{eqnarray}
\sum_{\boldsymbol{n}}\chi_{n_{t}}^{\uparrow_b}(\boldsymbol{n})
= \sum_{\boldsymbol{n}} \chi_{n_{t+1}}^{\uparrow_b}(\boldsymbol{n}) = 1.
\end{eqnarray}

We now treat the impurities as fixed worldlines and integrate them out from the lattice action. 
This leads to the ``reduced'' transfer matrix $\slashed{M}\K{n_t}$ given in Eq.~(\ref{eq: matel}), 
which is acting on the wave functions of the majority particles only,
\begin{align}
\begin{split}\label{eq: matel}
&\matrixel{\chi_{n_{t+1}}^{\downarrow},\chi_{n_{t+1}}^{\uparrow_a},\chi_{n_{t+1}}^{\uparrow_b}}
{   \hat{M}  }
{\chi_{n_{t}}^{\downarrow},\chi_{n_{t}}^{\uparrow_a},\chi_{n_{t}}^{\uparrow_b}}=\\
&\prod_{\vn}\K{\Ke{\frac{\overrightarrow{\partial}}{\partial\theta_\downarrow^*\K{n_{t},\vn}}}^{\chi_{n_{t+1}}^{\downarrow}\K{\vn}}}
X^{\slashed{\uparrow}\slashed{\uparrow}}\K{\vn}\slashed{M}\K{n_t}\\
\times&\left.\prod_{\vn'}\K{\Ke{\frac{\overleftarrow{\partial}}{\partial\theta_{\downarrow}\K{n_{t},\vn'}}}^{\chi_{n_{t}}^{\downarrow}\K{\vn'}}}\right |_{
\theta_\downarrow=\theta^*_\downarrow=0},
\end{split}
\end{align}
where $$X^{\slashed{\uparrow}\slashed{\uparrow}}=\prod_{\vn} \exp\K{\theta^*_\downarrow\Ke{n_t,\vn}\theta_\downarrow \Ke{n_t,\vn}}.$$ 
We consider the reduced transfer matrix between two successive Euclidean time steps $n_t$ and $n_{t+1}$, and this corresponds to three different cases. 

The first case is when both impurities hop from lattice 
sites $(\vn,\vn')$ to nearest neighbor sites $(\vn\pm\hat{l},\vn'\pm\hat{l}')$. The amplitude for 
one worldline to hop to a neighboring lattice site is
\begin{equation}
\label{eqn:Wh}  
W_h=\frac{\alpha}{2m},
\end{equation}
while the amplitude to stay on the same lattice site is given by
\begin{equation}
W_s=1-6W_h.
\end{equation}
When this is applied to Eq.~(\ref{eq: matel}), the reduced transfer matrix reads
\begin{align}
\slashed{M}_{\vn'\pm\hat{l}',\vn'}^{\vn\pm\hat{l}, \vn}&=W_h^2e^{-\alpha H_0^\downarrow},
\label{eqn:reducedM-case1}
\end{align}
where the upper set of indices of $\slashed{M}$ refers to the new and old position of worldline $a$
while the lower set of indices refers to the positions of worldline $b$.  

The second case results when one of the impurities (either
worldline $a$ or worldline $b$) hops to some nearest neighbor site while the other one remains stationary. 
The corresponding transfer matrix elements for  $(\vn,\vn') \to (\vn,\vn'\pm\hat{l}')$ and $(\vn,\vn') \to (\vn\pm\hat{l},\vn)$ are 
equivalent, 
\begin{align}
\slashed{M}_{\vn\vn}^{\vn'\pm\hat{l}, \vn'}=\slashed{M}_{\vn'\pm\hat{l},\vn'}^{\vn \vn}.
\end{align}
We can then write
\begin{align}
\slashed{M}_{\vn\vn}^{\vn'\pm\hat{l}, \vn'}
=W_sW_he^{-\alpha H_0^\downarrow-\frac{\alpha C_{IB}\theta^*_\downarrow\K{n_t,\vn}\theta_\downarrow\K{n_t,\vn}}{W_s}}\;.
\label{eqn:reducedM-case2}
\end{align}
From this result, we find that only the worldline that does not hop between time steps $n_t$ and $n_{t+1}$ can 
interact with the background particles.

The third and last case is where both worldlines remain stationary between time steps $n_t$ and $n_{t+1}$,  $(\vn,\vn') \to (\vn,\vn')$, 
and the reduced transfer matrix is
\begin{align}
\slashed{M}_{\vn\vn}^{\vn'\vn'} = & \: W_s^2e^{-\alpha H_0^\downarrow} \exp \left[-\frac{\delta_{\vn,\vn'}\alpha C_{II}}{W_s^2}\right.\nonumber\\
&+\left(1-{\alpha C_{IB}\frac{\theta^*_\downarrow\K{n_t,\vn}\theta_\downarrow\K{n_t,\vn}
        }{W_s}}\right)\nonumber\\
&\times\left.\left({1-\alpha C_{IB}\frac{\theta^*_\downarrow\K{n_t,\vn'}\theta_\downarrow\K{n_t,\vn'}
        }{W_s}}\right)\right]\,.
\label{eqn:reducedM-case3}
\end{align}
As can be seen from Eq.~(\ref{eqn:reducedM-case3}), the stationary worldlines do interact with the background particles. 
If the worldlines are occupying the same spatial lattice site between time steps $n_t$ and $n_{t+1}$, then they can also can interact with each other. 
Therefore, by considering all possible configurations of the worldlines, we obtain the reduced transfer matrix elements to be used in our MC calculations. 
In Eqs.~(\ref{eqn:reducedM-case1}-\ref{eqn:reducedM-case3}) all expressions multiplied with $e^{-\alpha H_0^\downarrow}$ can 
be viewed as local auxiliary fields felt by the background particles. 

\begin{figure}[t!]
\begin{center}
\includegraphics[width=0.45\textwidth]{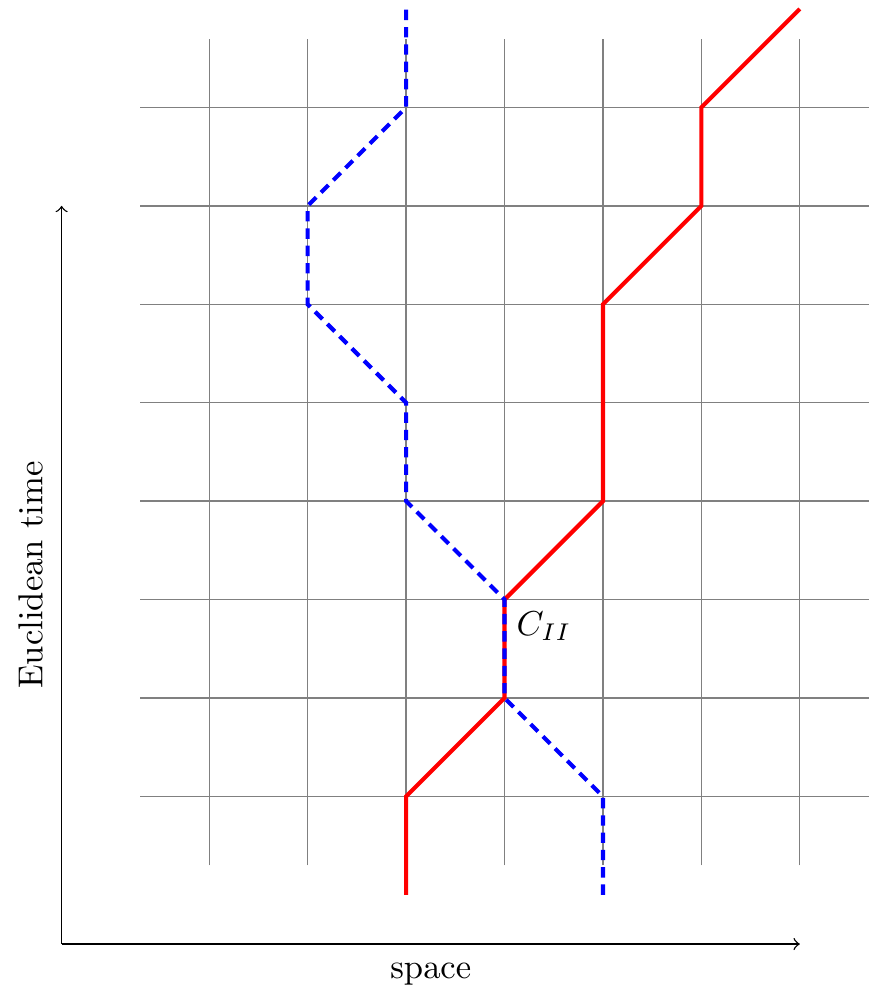}
\caption{Illustration of a representative worldline configuration. When the two worldlines (solid and dashed) 
stay at the same lattice site for one Euclidean time step, they interact with the interaction strength $C_{II}$.  \label{fig: wlgrid}}
\end{center}
\end{figure}

From the scenarios discussed above, we obtain the reduced transfer matrix operators
\begin{align}
\label{operators}
\slashed{M}_{\vn'\pm\hat{l}',\vn'}^{\vn\pm\hat{l}, \vn} &=  \: W_h^2:e^{-\alpha H_0^\downarrow}:,\\
\hat{\slashed{M}}_{\vn\vn}^{\vn'\pm\hat{l}, \vn'} & = \: W_h W_s 
:e^{-\alpha H_0^\downarrow-\frac{\alpha C_{IB} \, \rho_\downarrow\K{\vn}}{W_s}}:,\\
\hat{\slashed{M}}_{\vn\vn}^{\vn'\vn'}
& =  \: W_s^2 :e^{-\alpha H_0^\downarrow}\exp\bigg[\frac{-\delta_{\vn,\vn'}\alpha C_{II}}{W_s^2} \nonumber\\
&-\frac{\alpha \, C_{IB} \, \rho_\downarrow\K{\vn}}{W_s}
-\frac{\alpha C_{IB}\, \rho_\downarrow\K{\vn'}
}{W_s} + \mathcal{O}(\alpha^2)\bigg]:\,.
\end{align} 
These operators are only acting on the spin-down background particles.
One  example configuration for two worldlines is depicted in  Fig.~\ref{fig: wlgrid}.
In comparison with the one impurity case, the evaluation of Eq.~\eqref{eq: matel} gives rise to an additional term associated with the impurity-impurity interaction term $C_{II}$. 
%%%%%%%%%%%%%%%%%%%%%%%%%%
In order to illustrate this behaviour we consider the case where both worldlines stay
on the same lattice side ($\vn=\vn'$). After expanding the exponentials, the expression for the 
matrix element reads
\begin{align}
%\begin{split}
& \matrixel{\chi_{n_{t+1}}^{\downarrow},\chi_{n_{t+1}}^{\uparrow_a},\chi_{n_{t+1}}^{\uparrow_b}}
{ \hat{M} }
{\chi_{n_{t}}^{\downarrow},\chi_{n_{t}}^{\uparrow_a},\chi_{n_{t}}^{\uparrow_b}}
\nonumber \\
& = W_s^2 \prod_{\vn}\K{\Ke{\frac{\overrightarrow{\partial}}{\partial\theta_\downarrow^*\K{n_{t},\vn}}}^{\chi_{n_{t+1}}^{\downarrow}\K{\vn}}}
X^{\slashed{\uparrow}\slashed{\uparrow}}\K{\vn}
\nonumber \\
& \times :\K{1-\alpha H_0^\downarrow} \!\!
\bigg[1-\frac{\alpha C_{II}}{W_s^2}-2\alpha C_{IB}\frac{\theta^*_\downarrow\K{n_t,\vn}\theta_\downarrow\K{n_t,\vn}}{W_s}\bigg]:
\nonumber \\
& \times\left.\prod_{\vn'}\K{\Ke{\frac{\overleftarrow{\partial}}{\partial\theta_{\downarrow}\K{n_{t},\vn'}}}^{\chi_{n_{t}}^{\downarrow}\K{\vn'}}}\right |_{\theta^{\K{*}}_\downarrow=0}.
%\end{split}
\end{align}
In contrast to the one-impurity case, the term proportional to $C_{II}$ is an interacting term that is not
proportional to the background field $\theta_\downarrow$. It can be therefore combined with fields coming
from $H_0^\downarrow$ or $X^{\slashed{\uparrow}\slashed{\uparrow}}$. This leads to a contribution of the form
\begin{align}
\begin{split}
&\matrixel{\chi_{n_{t+1}}^{\downarrow},\chi_{n_{t+1}}^{\uparrow_a},\chi_{n_{t+1}}^{\uparrow_b}}
{ \hat{M} }
{\chi_{n_{t}}^{\downarrow},\chi_{n_{t}}^{\uparrow_a},\chi_{n_{t}}^{\uparrow_b}}\\
&= W_s^2\K{W_s-\alpha\frac{C_{II}}{W_s}-2\alpha \frac{C_{IB}}{W_s}}.
\end{split}
\end{align}
%%%%%%%%%%%%%%%%%%%%%%%%%%%
%---------------------------------------------------------------------------------
%-----------------------------------------------------------------------------------

%%%%%%%%%%%%%%%%%%%%%%%%%%%%%%%%%%%%%%%%%%%%%%

\section{Results \label{sec:res}}

In our numerical ILMC calculations, we set the spatial lattice spacing to $1/a = 100$~MeV, and the temporal lattice spacing to 
$1/a_t = 300$~MeV, and thus $\alpha = a_t/a = 1/3$. We use a periodic cubic lattice with length $L = 10$ (in units of $a$).
We use the average nucleon mass $m=938.92$~MeV for the mass of the fermions. The impurity-background interaction $C_{IB}$ is taken to be 
attractive and tuned to give a natural-sized scattering length, in this case~$3$~fm.
This is comparable to the hyperon-nucleon interaction scattering length, see e.g. the recent chiral EFT work in Ref.~\cite{Haidenbauer:2019boi}.
As a check and benchmark of our formalism, we also calculate the triton binding energy for a Wigner SU(4) symmetric interaction adopting 
the AFMC code used in Ref.~\cite{Lu:2018bat}. For further details, see the Appendix.

\subsection{Repulsive impurity-impurity interaction}

\begin{figure}[tp]
\begin{center}
\includegraphics[width=0.50\textwidth]{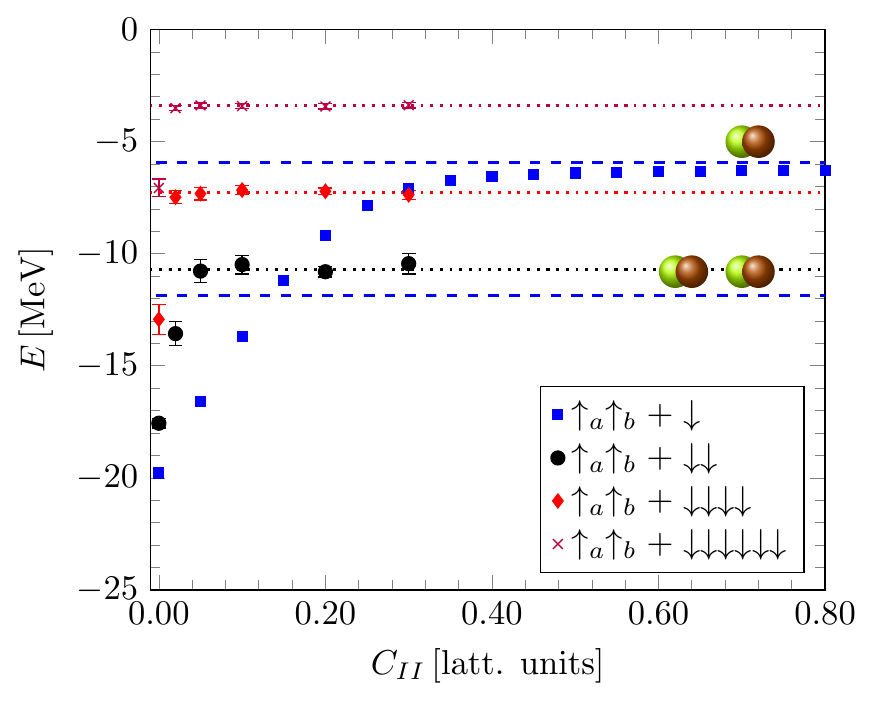}
\caption{Energy for $N_{BG}=1,2,4$ and $6$ background particles. The dotted lines are extrapolations
to infinite repulsion, while the dashed lines represent the one- and two-dimer thresholds.
\label{fig:repulsive}}
\end{center}
\end{figure}

We first consider the case where impurity-impurity interaction
is repulsive and vary the interaction strength $C_{II}$ to investigate the
formation or disintegration of $N$-body bound states for $N=3,4,6,8$. In Fig.~\ref{fig:repulsive}, we show the ILMC results for a periodic box with length $L = 10$ lattice units. The dashed lines in Fig.~\ref{fig:repulsive} represent the one- and two-dimer thresholds at infinite volume.

We note that the $N=3$ ground state lies below the one-dimer threshold even for very repulsive $C_{II}$, suggesting that the trimer bound state remains bound even for strongly repulsive impurity-impurity interactions.  Meanwhile, the $N=4$ ground state quickly moves to a position slightly above the two-dimer threshold for repulsive $C_{II}$, indicating that the ground state of the four-body system consists of two dimers.  A finite-volume analysis of the offset relative to the two-dimer threshold would extract the properties of the dimer-dimer interaction.  

Such re-configuration of the two impurity particles is rather interesting. The two impurities are bound together into a trimer when there is only one background particle.  But when there are two background particles, the two impurities prefer to separate and form two dimers. For $N=6$ and $N=8$ we find that the extra background fermions do not provide additional binding energy.  Instead, they are filling the Fermi sea of our periodic box.

%---------------------------------------------------------------------------------
%-----------------------------------------------------------------------------------

%%%%%%%%%%%%%%%%%%%%%%%%%%%%%%%%%%%%%%%%%%%%%%

\subsection{Attractive impurity-impurity interaction}

\begin{figure}[tp]
\begin{center}
\includegraphics[width=0.50\textwidth]{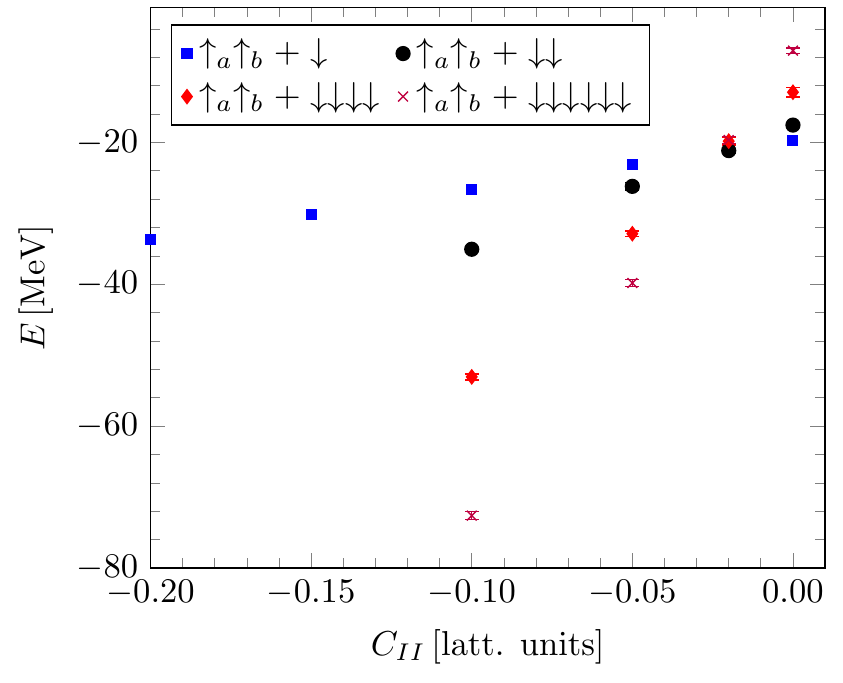}
\caption{Energy for attractive impurity impurity interaction for different numbers of background particles $N_{BG}=1,2,4$ and $6$.
\label{fig:attractive}}
\end{center}
\end{figure}

In similar fashion as in the repulsive case, we perform an analysis for the
attractive impurity-impurity interaction. We vary the coupling constant $C_{II}$ in order to
study the appearance of $N$-body states for $N=3,4,6,8$.  The ILMC results are shown in Fig.~\ref{fig:attractive}.  We see that the $N=4$ ground state energy drops below the $N=3$ ground state at around $C_{II}\approx 0.02$.  In fact, $C_{II}\approx 0.02$ is also where the $N=6$ ground state drops below the $N=4$ ground state and where the $N=8$ ground state drops below the $N=6$ ground state.  This crossing of several energy levels at the same point is reminiscent of a quantum phase transition.  In this case, however, we are considering a finite system with relatively few particles.  This rich phase structure will be explored in a future publication.
We note another recent study that considered the properties of two heavy
impurities in a Fermi bath which also found similar bound states~\cite{Sighinolfi:2021opb}.
 
While the system we study here does not directly correspond to double $\Lambda$
hypernuclei, there are some interesting parallels worth noting.  The attractive
$\Lambda\Lambda$ interaction is not strong enough to to produce a  $\Lambda\Lambda$
bound state, see e.g. Ref.~\cite{Gal:2016boi}.  However, the attraction is
sufficient to help in the binding of hypernuclei.

%
%
%---------------------------------------------------------------------------------
%-----------------------------------------------------------------------------------

%%%%%%%%%%%%%%%%%%%%%%%%%%%%%%%%%%%%%%%%%%%%%%

\section{Discussion \label{sec:disc}}

We have shown, as a proof of principle, that a two-impurity Monte Carlo approach
is suitable to describe systems with more than one impurity, immersed in a sea of majority
particles. While the majority particles have no direct interactions with each other, an 
effective interaction is mediated by interactions with the impurities.  Moreover, we 
have observed the expected disappearance of complex bound states into dimers, for 
a sufficiently strong and repulsive impurity-impurity interaction. For an attractive impurity-impurity interaction, 
we have observed the expected formation of bound states when we increase the interaction
strength, which we expect to give rise to a rich and complex phase diagram.

We recall that the scenario studied here is formulated in terms of an impurity-impurity
interaction similar to a hyperon-hyperon interaction, which allows for a straightforward application
to $\Lambda\Lambda$ nuclei.
In addition, it would be worthwhile to study the emergence of few-body states in the context of
ultra-cold gases, as particle mixtures could be treated as impurities. In such a case,
the couplings can be tuned such that one is at (or close to) unitarity, in order to map out the
universal features of such systems, see e.g. Ref.~\cite{Naidon:2016dpf}.

%%%%%%%%%%%%%%%%%%%%%%%%%%%%%%%%%%%%%%%%%%%%%%%%%%%%%%%%%%%%%%%%%%%%%

\section*{Acknowledgments}

We thank 
Johann Haidenbauer, Andreas Nogga and Dillon Frame
for useful discussions.
This work was supported in part by the European
Research Council (ERC) under the European Union's Horizon 2020 research
and innovation programme (grant agreement No. 101018170),
by DFG and NSFC through funds provided to the
Sino-German CRC 110 ``Symmetries and the Emergence of Structure in QCD" (NSFC
Grant No.~11621131001, DFG Grant No.~TRR110).
The work of UGM was supported in part by VolkswagenStiftung (Grant no. 93562)
and by the CAS President's International
Fellowship Initiative (PIFI) (Grant No.~2018DM0034).
The work of DL is supported in part by the U.S. Department of Energy (Grant 
No. DE-SC0018638) and the Nuclear Computational Low-Energy Initiative (NUCLEI) SciDAC project.
The authors gratefully acknowledge the Gauss Centre for Supercomputing e.V. (www.gauss-centre.eu) 
for funding this project by providing computing time on the GCS Supercomputer JUWELS 
at J\"ulich Supercomputing Centre (JSC).

%%%%%%%%%%%%%%%%%%%%%%%%%%%%%%%%%%%%%%%%%%%%%%%%%%%%%%%%%%%%%%%%%%%%%
\appendix
\section{Benchmark Calculation \label{sec: app}}
\begin{figure}[tp]
\begin{center}
\includegraphics[width=0.50\textwidth]{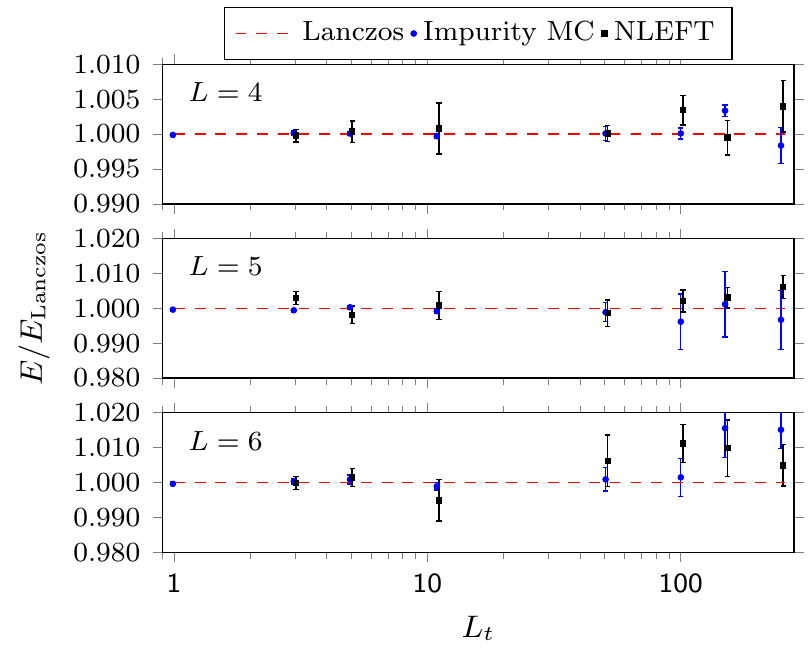}
\caption{Benchmark calculation for the two-impurity MC algorithm. The interaction is fixed in such a way to reproduce the 
  triton binding energy. The red dashed lines represent the exact Lanczos calculation results, which are
  used for the normalization,  the blue circles the two-impurity results
and the black squares the result of the NLEFT code. The comparison was performed for different box sizes $L=4,5$ and $6$.
\label{fig: Benchmark}}
\end{center}
\end{figure}

In order to benchmark our calculation, we compare our two-impurity MC calculation with two
established methods. The first one is a three-body Lanczos code that solves the problem iteratively (without MC simulation). The 
second one is the well-established NLEFT code~\cite{Lu:2018bat}. For this benchmark, we switch off all 
additional features of the NLEFT code (such as operator smearing and pion exchange) and use a SU(4) symmetric force which is tuned
to reproduce the triton binding energy at $L=35$. We then perform such calculations for $L=4,5,6$ to check
whether the three equivalent formulations of the same physical problem agree. As seen in Fig.~\ref{fig: Benchmark}, 
this check produces consistent results within statistical errors. Note also that we undertook no computational effort 
to reduce the errors at large Euclidean times. We therefore conclude that two-impurity Monte Carlo is well suited to 
address hypernuclear problems in the future.  

%%%%%%%%%%%%%%%%%%%%%%%%%%%%%%%%%%%%%%%%%%%%%%
%---------------------------------------------------------------------------------
%-----------------------------------------------------------------------------------

\end{document}